\documentclass[useAMS,usenatbib]{mn2e}

\usepackage{graphicx}

\title[Biases in samples of binary stars]{Twins like to be seen: Observational biases affecting spectroscopically selected binary stars}
\author[A. G. Cantrell and T. J. Dougan]{A. G. Cantrell\thanks{E-mail:
acantrell@blakeschool.org} and T. J.
Dougan\\
The Blake School, Minneapolis, MN, United States}
\begin{document}

\date{Accepted 2014 September 9.  Received 2014 September 7; in original form 2014 June 12}

\pagerange{\pageref{firstpage}--\pageref{lastpage}} \pubyear{2014}

\maketitle

\label{firstpage}

\begin{abstract}
Massive binary stars undergo qualitatively different evolution when the two components are similar in mass (`twins'), and the abundance of twin binaries is therefore important to understanding a wide range of astrophysical phenomena.  We reconsider the results of Pinsonneault \& Stanek (2006), who argue that a large proportion of binary stars  have nearly equal-mass components; we find that their data imply a relatively small number of such `twins.'  We argue that samples of double-lined spectroscopic binaries are biased towards systems with nearly equal-brightness components.  We present a Monte-Carlo model of this bias, which simultaneously explains  the abundance of twins in the unevolved binaries of Pinsonneault \& Stanek (2006), and the lack of twins in their evolved systems.  After accounting for the bias, we find that their observed mass ratios may be consistent with a variety of intrinsic distributions, including either a flat distribution or a Salpeter distribution.   We conclude that the observed overabundance of twins in Pinsonneault \& Stanek (2006) does not reflect the true population of binaries, and we briefly discuss the astrophysical implications of the lack of twins.
\end{abstract}


\section{Introduction}
Binary stars are common in the universe and their evolution is important to many astrophysical phenomena including supernova rates, enrichment of the interstellar medium, and the formation of X-ray binaries and blue stragglers.  Pinsonneault \& Stanek (2006, hereafter PS06) discussed the importance of the distribution of binary mass ratios to understanding binary evolution: binaries with similar-mass components (twins) undergo qualitatively different evolution than binaries whose two components are quite different in mass.  This affects all aspects of binary evolution, and is of particular interest to the formation of double compact objects, which are potential sources of gravitational waves (PS06).  The size of the twin population is therefore important to understanding and modeling a wide range of astrophysical phenomena; see PS06 for an overview of the important role of twins.

PS06 studied the distribution of mass ratios ($q=M_2/M_1\leq1$) of OB binaries in the Small Magellanic Cloud (SMC), using the sample of eclipsing binaries published in Harries et al. (2003, hereafter HHH03) and Hilditch et al. (2005, hereafter HHH05).  PS06 divides this sample into detached and semidetached binaries and finds a very large proportion of twins among detached binaries: they estimate that 45\% of detached binaries have $q>0.95$.  By contrast, they find that the semidetached systems have  a low rate of twins, with a distribution of $q$ peaked near 0.6.  They argue that the difference between the two samples cannot be due to an observational bias and that the detached sample reflects the true proportion of twins among detached binaries in the SMC.

PS06 argued that the twin population is a dominant feature, not just of their sample, but of previously studied binary samples.  They specifically claim that a similar twin population is present in the samples of  Lucy \& Ricco (1979, henceforth LR79), Tokovinin (2000, henceforth T00), and  Halbwachs et al. (2003, henceforth HMUA03).  Indeed, LR79 and T00 find twin populations similar to that in PS06, as does Lucy (2006, henceforth L06); HMUA03  also finds an excess of twins, though their results are not directly comparable due to a different selection of primary masses and the inclusion of single-lined spectroscopic binaries (SB1s) in their sample.  HMUA03 shows that the SB1s have systematically lower mass ratios than the double-lined spectroscopic binaries (SB2s), leading to a lower number of twins in their overall population.

Since SB1s have systematically lower $q$ than SB2s, samples consisting entirely of SB2s (including LR79, T00, L06 and PS06) overestimate the true size of the twin population.  Although L06 makes a compelling case for the statistical significance of the twin population, they do not consider the bias introduced by using a sample consisting only of SB2s.  Raghavan et al. (2010) did a comprehensive multiplicity study of nearby stars and found a smaller twin population than appears in studies of SB2s; though they find evidence that this population increases at short orbital periods, their sample at these periods is extremely small.  Sana et al. (2012, henceforth S12) observed a sample of OB stars and performed a statistical analysis of the full population (SB2s, SB1s and single stars); they find that the underlying distribution of $q$ is close to flat and does not show clear evidence for a twin population.  

In this paper, we argue that samples of SB2s are biased in favor of binaries whose components have similar brightnesses, rather than similar masses {\it per se}.  We show that the full sample of PS06 -- detached and semidetached -- consists of binaries whose components are similar in brightness.  We present a Monte Carlo model showing that a consistent bias based on brightness ratio will affect the distribution of $q$ differently depending on the evolutionary status of the binaries.  Our model reproduces both the abundance of twins in the detached population and the lack of twins in the semidetached population.  We conclude that the data used by PS06 are consistent with a far smaller twin population than they find.

\section{Observational Bias in Spectroscopically Selected Samples of Binaries}

The OB binaries analyzed by PS06 were initially discovered as eclipsing binaries.  Given the quality of the photometry in HHH03 \& HHH05, non-twins ($q<0.95$) should be easily detected through their eclipses.  However, their sample is further selected by the fact that they only obtained orbital solutions for the SB2s.  Specifically, HHH03 and HHH05 obtained spectroscopy for 169 binaries, $\sim100$ of which received detailed followup, and only 50 of which (all SB2s) appear in the final sample used by PS06. Samples of SB2s are known to be biased towards twins: in samples including both SB1s and SB2s, the SB1s have systematically lower mass ratios.  For example, the SB1s in HMUA03 have $q<0.65$, while their SB2s have $q>0.65$ (their Fig. 6).  The 119 binaries not in their final sample are presumably SB1s and likely have lower mass ratios than the SB2s studied by PS06.

\subsection{SB2s and the role of brightness}

In general, we would expect a sample of SB2s to favor binaries in which both components contribute significantly to the observed spectrum.  Since HHH03 and HHH05 obtained spectroscopy at $4000-4800$\AA, this would bias their sample towards systems whose components have similar B-band magnitudes.  If such a bias is present, we would expect the SB2s in HHH05 and HHH03 to have brightness ratios near 1, even in cases where the mass ratio is significantly below 1.
 
Fig. 1 shows histograms of mass ratio and B-band brightness ratio ($2.51^{-|B_1-B_2|}$); following PS06, we split the sample into detached and semidetached binaries.  We computed absolute B-band magnitudes using the spectral types and radii listed in HHH03 and HHH05, together with the values of $F'_V$ and $B-V$ given in Popper (1980).  We exclude the two contact binaries, because the possibility of distortion makes this computation of B-band magnitude (from radius) questionable for these systems. Since the publication of PS06, de Mink et al. (2007) have proposed alternate semidetached solutions for two of the systems originally listed as detached. In order to obtain a direct contrast to the results of PS06, we will use the original solutions found in HHH03 and HHH05. 

\begin{figure}
\includegraphics[width=\columnwidth]{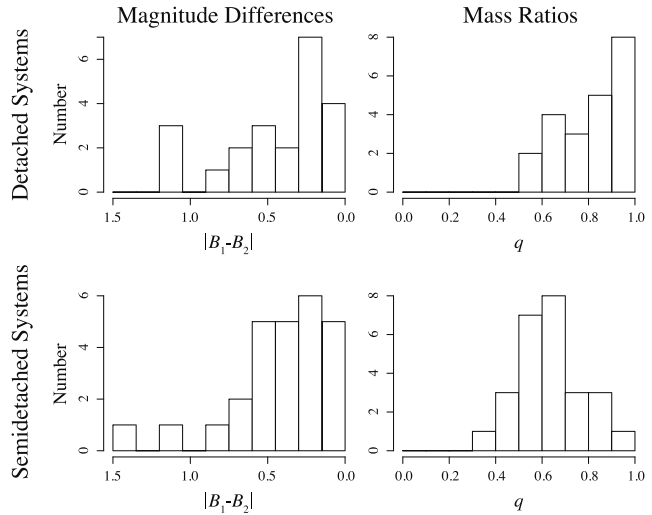}
 \caption{Histograms of B-band brightness ratio and $q$ for 48 detached and semidetached binaries in PS06. Although the two populations show very different distributions of $q$, the distributions of brightness ratio are similar: A K-S test comparing the two distributions of brightness ratio finds that they could be drawn from the same distribution with probability $p=0.51$.  This consistency indicates the possibility of a bias which favors similar-brightness binaries throughout the PS06 dataset.  \label{fig1}}
\end{figure}

Fig. 1 shows that the detached and semidetached samples of PS06 have similar B-band brightness ratios, despite their different distributions of $q$. This impression is confirmed by Kolmogorov-Smirnov (K-S) tests: A K-S test comparing the detached and semidetached distributions of $q$ gives $p=0.002$, whereas a K-S test comparing the two populations' brightness ratios gives $p=0.51$.  PS06 states that the inconsistent distributions of $q$ rule out the possibility of a single bias explaining both populations.  However, the consistent distributions of $|B_1-B_2|$ actually support the existence of a single bias, favoring similar-brightness binaries consistently in both samples. 

To understand the relationship between the distributions of mass ratio and brightness ratio, we show in Fig. 2 the relationship between absolute B-band magnitude (M$_B$) and mass ($M$) for the individual stars in the detached and semidetached binaries of PS06. As one would expect, the detached systems generally feature unevolved stars, for which mass and radius are strongly correlated.  The semidetached systems, by contrast, feature many evolved components, for which mass and radius are weakly correlated.  Equal-brightness binaries in the detached sample will therefore have similar masses, but this is not necessarily the case for the semidetached systems.   The presence of twins in the detached sample -- and the lack of twins in the semidetached sample -- can therefore be explained by a consistent bias and  the differing evolutionary status of the two samples.

\begin{figure}
\includegraphics[width=\columnwidth]{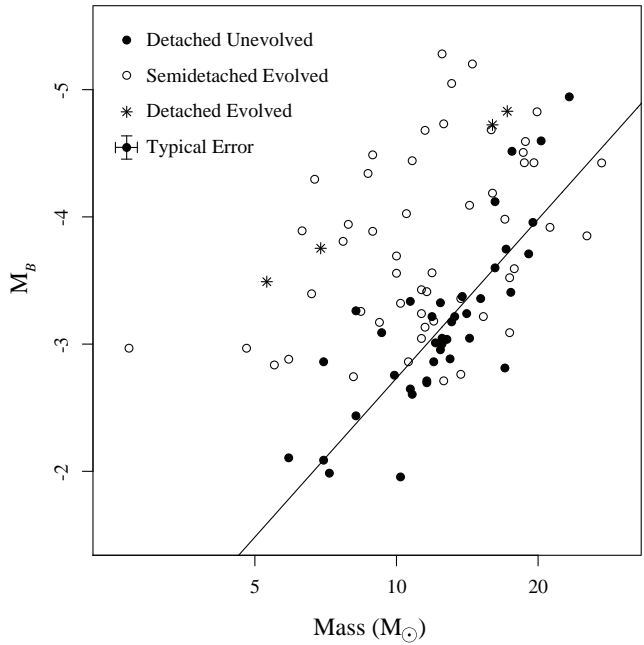}
\caption{Plot of absolute B-band magnitude and $\log(M)$ for 96 stars (primaries and secondaries) in the detached and semidetached binaries of PS06. Most of the stars in detached systems (solid circles) lie on the main sequence and show a strong correlation between mass and M$_B$.  Only four stars in two detached systems (asterisks) show clear evolution; these are classified as evolved in this paper. The stars in semidetached systems (open circles) are more evolved and thus have a weaker relationship between mass and M$_B$.  A bias favoring equal-brightness systems will have different effects on the observed distribution of $q$, depending on the relationship between mass and brightness; this is consistent with the discrepant distributions of $q$ shown in Fig 1.  \label{fig2}}
\end{figure}

If the abundance of similar-brightness binaries in PS06 is not due to a bias, then their sample must be representative of the actual population in the SMC.  In this case, binaries in the SMC must consistently maintain high brightness ratios at various stages of evolution.  In general, the low-$q$ semidetached systems of PS06 have a secondary (lower-mass) star which is more evolved than the primary: this is necessary in order to have binaries in which the secondary is as bright as the higher-mass primary.  For example, OGLE 09064499 has 2.7 and 8.4 M{$_\odot$} components, but the 2.7 M{$_\odot$} star has evolved to the point that the components are nearly equal in brightness ($2.51^{-|B_1-B_2|}=0.97\pm0.17$).  It is difficult to explain how this system will maintain a brightness ratio as high as those studied by PS06: Does such a binary form only when both stars are at an appropriate stage of evolution, then disband when either component evolves to outshine the other?  Or does mass transfer between the two stars occur in a way that maintains high luminosity ratios at all stages of evolution, transferring just enough to counteract evolution at every stage?  A detailed physical model would be necessary to make either of these explanations credible; absent such a model, the abundance of high brightness-ratio systems is best understood as the result of observational bias.

\subsection{A model for the observational bias}

We now use a Monte Carlo model to explore how a brightness ratio bias could affect evolved and unevolved populations of binaries differently. In comparing our model to the mass ratios of PS06, we divide the observed sample into a group of binaries in which both components are unevolved, and a group in which at least one star is evolved.  There are two detached systems with evolved components (asterisks in Fig. 2); other than these two systems, our unevolved and evolved samples are the same as the detached and semidetached samples studied by PS06.  We will show that a relatively simple model can explain the inconsistent mass ratio distributions of the evolved and unevolved samples.

Our Monte Carlo model begins with two populations of 100 million binaries each.  In both populations, single stars are given masses drawn from a Salpeter initial mass function (IMF) between 0.3 M$_\odot$ (low enough that our simulated selection effects eliminate all stars of this mass) and 30 M$_\odot$ (the highest mass present in PS06). The brightness of each single star is then computed either according to a fixed relationship (unevolved sample) or with some variation added to this relationship (evolved sample).  Finally, single stars are paired at random into binaries, and both populations are subjected to identical cuts simulating the observational biases present in PS06.  The only difference between the two samples is the relationship between the simulated masses and brightnesses of the single stars being observed.

For stars in the simulated unevolved sample, we compute  B-band magnitude directly from each star's mass.  Applying a linear regression to the unevolved stars (closed circles) in Fig. 2 gives M$_B=1.42-4.15\log(M)$; we use this formula to compute B-band magnitudes for the individual stars in the simulated unevolved sample.   Since our goal is specifically to model the stars observed by PS06 and the biases which may affect them, we derive our mass-brightness relationship from their sample rather than using a more theoretical model.

\begin{figure*}
\includegraphics[width=\textwidth]{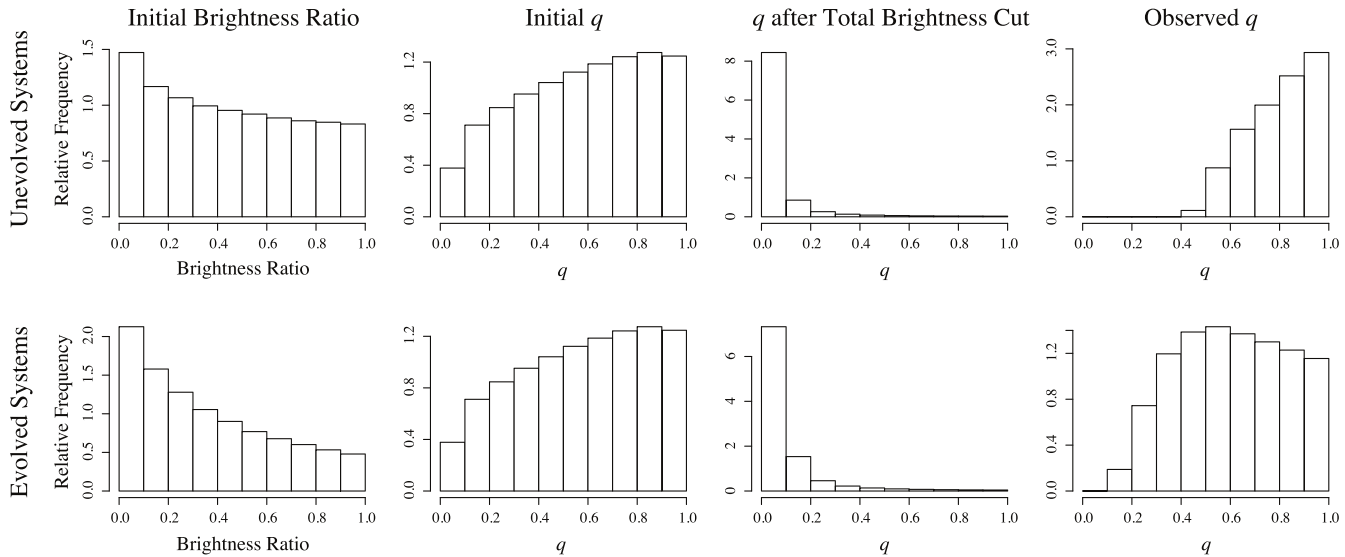}
\caption{Histograms showing the progression of Monte Carlo models of evolved and unevolved binary populations. The first two columns show the initial distributions of brightness ratios and mass ratios. The third column shows the distribution of mass ratios after a cutoff on total brightness, and the fourth column shows the distributions after an additional cut based on brightness ratio.  Both populations feature random pairings of single stars with masses drawn from a Salpeter IMF, and both are subjected to identical biases.  The only difference between the evolved and unevolved models is the algorithm for assigning individual stars a B-band magnitude.  Our model qualitatively reproduces the distributions of PS06, both the abundance of twins in the unevovled sample and the realtive lack of twins in the evolved sample. \label{fig3}}
\end{figure*}

For stars in the simulated evolved sample, each star's B-band magnitude is computed as the main sequence brightness given above ($B_{ms}=1.42-4.15\log(M)$) plus a randomly chosen offset representing the star's evolutionary status.  We start by assigning each star a random offset from the main sequence ($B-B_{ms}$), drawn from the distribution of $B-B_{ms}$ in the observed evolved sample of PS06.  After applying our selection effects (described below), we adjust the intrinsic distribution of $B-B_{ms}$ so that the observed distribution more closely resembles that of the PS06 evolved systems.  By an iterative procedure, we find a distribution of $B-B_{ms}$ (dependent on the observational bias) which perfectly matches the observed brightness offsets from the main sequence.
 
  We have modeled our stars on the sample of PS06 only in determining the brightness of {\it single} stars: these single stars, whose masses came from a Salpeter distribution, are paired randomly into binaries. Our model stars are rather simplistic, but our goal is not to present a fully realistic model and to unambiguously detemine the unbiased population.  Rather, we wish to show that a bias based on brightness ratio can have different effects on evolved and unevolved samples of binaries, and that this bias gives a simple explanation for both populations of PS06.

To simulate observational bias, the evolved and unevolved samples are both subjected to a cutoff on total brightness and a cutoff on B-band brightness ratio (modeling the difficulty of detecting a system as an SB2).  HHH05 applied an explicit brightness limit, $B<16$.  For the distance and reddening to the SMC, this is roughly equivalent to M$_I<-2.75$ (see fig. 12 of HHH05).  To implement our cut on total brightness, we use Bessell \& Brett (1988) to compute M$_I$ from M$_B$, then eliminate systems with M$_I>-2.75$.

We model the cutoff on brightness ratio not as a sharp cut, but rather as an increasing function of brightness ratio.  Whether a binary is observed as an SB2 probably depends on many things aside from brightness ratio, e.g. orbital velocity and inclination.  Perhaps a binary with brightness ratio $0.28$ will be an SB2 only if these other factors are ideally favorable, while systems with higher brightness ratios have more leeway.  Since 0.27 is the lowest brightness ratio present in PS06, we eliminate all simulated binaries with brightness ratio ($2.51^{-|B_2-B_1|}$) less than 0.27.  We then assume that the probability of observing a given binary is a linear function of brightness ratio between 0.27 and 1.  We will explore nonlinear models in Section 3, but the linear model is sufficient to demonstrate the essential properties of our bias.

\subsection{Results}

The results of our Monte Carlo model are shown in Fig. 3.  For the evolved and unevolved samples, we show the simulated distribution of brightness ratio and $q$ before selection effects, and the distribution of $q$ after each selection is applied.  Before selection effects are applied, the evolved and unevolved distributions of $q$ are identical (random draws from a Salpeter distribution).  The cut on total brightness favors low-$q$ systems in both the evolved and unevolved samples.  After the brightness ratio cut, the evolved and unevolved populations become distinct in just the way apparent in PS06: the unevolved sample is peaked at $q=1$, while the evolved sample is peaked at $q\sim0.55$.

The distribution of $q$ given by random draws from a Salpeter mass function is $p(q)\propto q^{0.35}$ (Tout 1991), shown in the second column of Fig. 3.  The cut on total brightness strongly favors low-$q$ systems; when mass and radius are correlated, this produces the Salpeter relative mass function used by PS06, $p(q)\propto q^{-2.35}$.  However, in the evolved sample the distribution of observed mass ratios is less peaked than $p(q)\propto q^{-2.35}$, due to the presence of luminous, low-mass stars. Despite the strong effect of the cutoff on total brightness, the brightness ratio cutoff eliminates the peak at $q=0$ and dominates the shape of the observed distribution.

\section{Discussion}

The model presented in Section 2 qualitatively reproduces the discrepant populations found by PS06.  To quantitatively compare the observed and simulated populations, we must account for observational error.  When PS06 modeled the data, they convolved the simulated mass ratios with observational error ($\sigma(q)=0.07$).  However, this gives numerous binaries in which the mass ratio (with error) is greater than 1, i.e. the secondary appears more massive than the primary; PS06 does not say how they handled these systems.  Rather than taking this approach, we add error to the individual simulated masses ($\sigma(M)=5\%$, from HHH05) and define the  mass ratio (with error) as the smaller mass (with error) divided by the larger mass (with error) giving $q<1$.

In Fig. 4, we make a direct comparison between the data of PS06 and our simulated populations including error. The simulated unevolved sample is a good match for the sample of PS06: A K-S test comparing the two distributions gives $p=0.45$.  For the evolved systems, our simulation actually overpredicts the observed number of low-$q$ systems.  A K-S test comparing the simulated evolved systems to those in PS06  gives $p=0.02$, a difference which could easily be due to our simplified model of binary evolution.  

Our bias assumes that the probability of detecting a system is a linear function of brightness ratio.  Although this is a simplification, our results are not strongly dependent on this assumption: the evolved and unevolved samples of PS06 can be well-modeled using a wide range of nonlinear selections on brightness ratio.  For example, the evolved and unevolved systems of PS06 can be equally well modeled by (1) the linear cutoff described above, (2) a probability of detection which is 0 for brightness ratios below 0.5 and 1 above 0.5, and (3) A model in which the probability of detection has no lower cutoff and goes as the brightness ratio to the 3.3.  Models 1, 2, and 3 match the unevolved population with K-S tests $p=0.45$, $p=0.20$, and $p=0.61$, respectively, and the evolved population with $p=0.02$, $p=0.06$, and $p=0.04$, respectively. These models are also plotted in Fig. 4. Determining the true nature of the brightness ratio cut would require simulating binary spectra and determining which could be observed.  However, such details are not essential to our result: the evolved and unevolved samples of PS06 can be successfully modeled by a variety of cuts on brightness ratio.

After convolving with observational error, a K-S test comparing the detached binaries to the `twin' model of PS06 (and assuming no observational bias) gives $p=0.09$  \footnote{PS06 does not give a K-S value comparing their twin distribution to the observed one; this value was derived by replicating their work, but convolving with error as described in the text.  PS06 added error directly to $q$, which results in many mass ratios greater than 1.  We suspect that PS06 simply discarded these systems; doing so gives a much better K-S test ($p=0.55$).  However, doing so artificially removes many high-$q$ systems from the model and therefore inflates the apparent size of the twin population.}, compared to $p=0.45$ for our model.  Modeling the twin population as a bias therefore reproduces the data better than the model proposed by PS06, despite having one free parameter (the lower cutoff) rather than three (twin fraction and the lower cutoffs of two flat distributions).  In addition, our model provides a natural explanation for the lack of twins among evolved systems; this feature is not explained by PS06.  Finally, our model studies the consequences of a brightness ratio bias which is generally expected: PS06 acknowledged such a bias, but rejected the possibility that it could affect evolved and unevolved binaries differently.  We conclude that the mass ratios studied by PS06 are best understood as the result of an observational bias.

\begin{figure}
\includegraphics[width=\columnwidth]{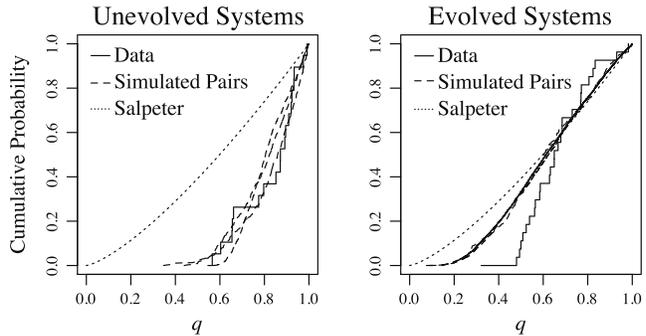}
\caption{Cumulative probability distributions of $q$ for the 48 unevovled and evolved systems of PS06 (solid lines), compared to the results of our Monte Carlo models of observational biases (dashed lines; three in each panel, corresponding to the models described in the text). For reference, the Salpeter mass ratio distribution, the intrinsic distribution of $q$ for the simulated samples (dotted line) is also provided. The dashed lines include convolution with observational error, as described in the text. Our model produces a good fit to the unevolved systems: a K-S test comparing our model to the data gives $p=0.45$.  Our model qualitatively reproduces difference between the evolved and unevolved systems, but significantly overestimates the size of this difference: it predicts even more low-$q$ evolved systems (and fewer evolved  twins) than are observed.   \label{fig4}}
\end{figure}

\subsection{Other intrinsic distributions of $q$}

S12 found that OB binaries come from a flat distribution ($q^\kappa$, $\kappa=-0.1\pm0.6$), in contrast with the $q^{-2.35}$ distribution predicted by a Salpeter mass function.  Adopting a flat distribution of OB binaries and applying our simulated bias produces a distribution of detached systems consistent with the data of PS06: a K-S test comparing them gives  $p=0.47$.  The simulated evolved systems also give similar results whether one starts with a Salpeter distribution or a flat distribution: $p=0.02$ and $p=0.03$ respectively.  The flat and Salpeter distributions are quite different, yet after modeling the bias both produce samples consistent with the unevolved systems of PS06 and qualitatively consistent with the evolved systems of PS06.  

The bias we model here is so strong that a range of intrinsic distributions give observed samples similar to those found by PS06.  As long as the initial distribution is roughly flat for $q \ga 0.7$, the observed distribution will be consistent with the data of PS06.  A significant over- or under-abundance of twins is the only sort of intrinsic distribution for which which the simulated observations will be inconsistent with the data of PS06.  In particular, adopting the PS06 twin population as the intrinsic distribution of $q$ (and applying our simulated bias) results in an observed sample which is a poor match for the data: a K-S test comparing them gives $p=0.004$.

If our bias is realistic, then a wide range of intrinsic distributions should result in similar observed distributions, in which case observed samples of SB2s should have very similar distributions of $q$ even if the underlying populations are different.  Indeed, the unevolved SB2s in HMUA03 (table 4 of their paper) have a $q$  distribution consistent with the unevolved systems of PS06: a K-S test comparing them gives $p=0.47$. These two populations (other than being SB2s) were selected in very different ways.  Whereas the binaries of PS06 were selected based on a brightness cut (effectively a minimum primary mass), HMUA03 selected systems with F7-K primaries (a narrow range of primary mass).  These two procedures should result in different intrinsic distributions of $q$.    The agreement between the two observed samples therefore reinforces the idea that the observed distributions are dominated by a consistent bias affecting samples of SB2s.

Although we conclude that the data of PS06 are best explained as the result of a bias, we caution against any effort to reverse this bias and thereby determine the underlying distribution.  The bias we propose is strong enough that it is impossible to reliably determine the underlying distribution given an observed sample of SB2s.    Rather, to study the true distribution of $q$, we recommend a study similar to S12 or Raghavan (2010), in which SB2s are part of a larger sample.  PS06 claims that their large twin population is supported by  Lucy \& Ricco (1979) and Tokovinin (2000).  However, both these papers considered unevolved SB2s (like the detached systems in PS06) and are therefore subject to the same biases.

\subsection{Comparison to PS06}

We have analyzed the dataset of PS06 and come to a conclusion strongly inconsistent with theirs.  We find that the true twin population may be negligible, whereas PS06 found $\sim45$\% of systems have $q>0.95$.  We now summarize the ways in which our analysis differs from theirs, and the contribution of these differences to the size of the twin population.  

(1) When we convolve our model with observational error, we add error to individual stars rather than to $q$.  As discussed in the introduction to section 3, convolving error directly with $q$ may cause some systems to be discarded and thereby inflate the implied size of the twin population. We redid the fit of PS06, modeling the detached systems as a flat distribution ($q>0.5$) and a twin population ($q>0.95$) but adding error to individual stars rather than to $q$.  The resulting best-fit twin fraction is then $29\pm7$\%.  Our method of convolution reduces the twin population by a third even before observational biases are considered.

(2) PS06 noted that if the lack of systems with $q<0.5$ is due to a bias, then this would effectively cut the number of twins in half.  Indeed, we believe that these systems are missed because of observational bias: the lowest brightness ratio observed in any system in PS06 (detached or semidetached) is 0.27.  For unevolved binaries (using our fit to Fig 2) this corresponds to $q\sim0.44$.  Attributing these missing systems to a bias reduces the 29\% twin fraction down to $16\pm3.5$\%.  

(3) The remaining difference between our twin fraction and that claimed by PS06, comes from modeling the brightness ratio bias not as a sharp cut but as graduated over a range of brightness ratio.  This graduated cutoff results in removing some systems with brightness ratio $>0.27$; it systematically removes relatively low-$q$ systems and (in our model) reduces the twin fraction to be consistent with the $7$\% resulting from random draws from a Salpeter distribution.

Of the three differences described above, (1) and (2) are quite robust. These two issues by themselves reduce the twin population from $45$\% to $16\pm3.5$\%.
Our graduated model of the bias (3) may be wrong in its details, but it is very likely right in its essence.  Even if the linear model is flawed, it is likely that the probability of detecting a binary as an SB2 increases over some range of brightness ratio, and that the true twin population is therefore below 16\%. 

\section{Conclusions}

We find that PS06 significantly overestimated the size of the twin population by employing a flawed method of error convolution and by neglecting the important role of observational bias.  After correcting for these two effects, the data of PS06 are consistent with either a Salpeter distribution of mass ratios, or with the flat distribution of S12; they do not provide a compelling case for a significant population of twins.  

In addition, we find that all binaries observed by PS06 have high brightness ratios, suggesting a consistent bias favoring these binaries.   We present a Monte Carlo model of this bias, which naturally explains both the abundance of twins in detached (unevolved) systems and the lack of twins in semidetached (evolved) systems.  Since this bias affects the distribution of $q$ differently in these two samples, the semidetached sample is not an `ideal control'  sample as was claimed by PS06.  

Compared to the model of PS06, modeling the twin population as the result of an observational bias has various advantages: (1) this bias is generally expected, and was acknowledged but not modeled by PS06, (2) our model gives a better fit to the mass ratios of the detached (unevolved) systems with fewer free parameters, and (3) our model qualitatively explains the semidetached (evolved) systems which are not addressed by PS06.  

The smaller twin population implied by our analysis has various astrophysical implications. Neutron star-black hole systems may be more abundant than neutron star-neutron star systems as gravitational wave sources, consistent with the non-twin population discussed by PS06 but contrary to their conclusion based on a large twin population.  In addition, binary mergers are likely to account for a relatively small fraction of blue stragglers, as coalescence is more likely in twin binaries (PS06).  A relatively small twin population also helps explain the abundance of black hole X-ray binaries, whose formation requires a relatively low-mass secondary (e.g. Kalogera \& Webbink 1998).  PS06 additionally discusses the important role of twins in forming type Ia supernovae and binary white dwarf systems.

The bias we propose is strong enough that the underlying distribution cannot be reliably recovered by correcting for the bias: a range of intrinsic distributions is consistent with the sample of PS06. This bias likely affects all samples of SB2s similarly. Although we cannot make a precise determination of the true size of the twin population, we find that the twin population ($q>0.95$) implied by the sample of PS06  is under 16\%.  The least certain aspects of our model may affect how far below 16\% the twin fraction is, but we strongly rule out a twin population of the size identified by PS06.  We conclude that a bias similar to the one discussed here has likely affected all observed samples of SB2s, calling into question the large twin populations found by PS06 and other papers based exclusively on SB2s, including T00, L06 and LR79. 

\section*{Acknowledgments}
 
We would like to thank Steve Kaback, Rand Harrington and The Blake School for their support of this work.  We thank J.L. Halbwachs, T. Maccarone and an anonymous referee for their comments on this paper.

\label{lastpage}

\end{document}